\documentclass[fleqn]{revtex4-2}

\usepackage{amssymb,amsmath}
\usepackage{graphicx}
\usepackage{verbatim}

\begin{document}

\title{Relativistic recoil corrections of order \boldmath$m(Z\alpha)^6(m/M)$: Hydrogen-like atom and hydrogen molecular ion.\\
}

\author{V.I.~Korobov}
\email{korobov@theor.jinr.ru}
\affiliation{Bogoliubov Laboratory of Theoretical Physics, Joint Institute for Nuclear Research, Dubna 141980, Russia}

\begin{abstract}
The finite operators are derived for the relativistic recoil corrections of order $m\alpha^6(m/M)$ to spin-averaged energy levels in hydrogenlike atoms and ions in two- and three-body formalisms beyond the adiabatic approximation. Results are presented in a form suitable for numerical evaluation.
\end{abstract}

\maketitle

\section{Introduction}

The important contribution in the calculations of the rovibrational state of the hydrogen molecular ions is the relativistic corrections of order $m\alpha^6$, so far these corrections have been performed in the adiabatic approximation \cite{Korobov08,Korobov21} and without taking into account recoil. In Ref.~\cite{Zhong18} general formulas for the $m\alpha^6$ and $m\alpha^6(m/M)$ order corrections were derived, but without elimination of singularities.

Recoil correction appears due to the finiteness of the nuclear mass. Since the electron-to-nucleus mass ratio, $m/M$, is small for hydrogen atom (or hydrogen molecular ion), it is convenient to expand the recoil corrections in terms of $m/M$ expansion. The study of mass recoil corrections has a long history: the Bethe-Salpeter equations were used in Ref.~\cite{Salpeter52}, then the modified Dirac equation was widely employed \cite{GrotchYennie}, and in recent years the expansion in powers of $m/M$ has been intensively developed \cite{Shabaev85,Yelkhovsky94,Pachucki95,Melnikov02}.

In our previous work \cite{Korobov25} we considered the leading $m\alpha^6$ order relativistic corrections in the expansion in $m/M$ and derived a set of finite (integrable) operators needed to numerically calculate this contribution. The goal of this paper is to continue our study and extend our approach to the next order $m\alpha^6(m/M)$ to obtain a new set of finite operators that are needed to compute the leading recoil contribution. For simplicity, we omit $Z$ from our notations of orders in the expansion in $(Z\alpha)^n$. We use the effective Hamiltonian approach derived from nonrelativistic quantum electrodynamics (NRQED).

In our derivation, we closely follow and compare our results with those based on the dimensionally regularized NRQED \cite{YPP16} obtained for the case of helium atom. This allows us to reliably check the final formulas in our approach.

The paper is organized as follows. In Sec. II, we formulate the problem for recoil corrections of order $m\alpha^6(m/M)$ for the case of hydrogenlike atoms. In Sec. III, we show how the divergent terms can be separated out from the regular part, and then in Sec. IV we define the high-energy contribution expressed as a $\delta$-function operator and show how it depends on the choice of regularization. In Sec. V we present our new results for obtaining the recoil correction of order $m\alpha^6(m/M)$ for the case of hydrogen molecular ions. In the Appendix we use the dimensionally regularized NRQED to rigorously derive from scratch the relativistic recoil corrections of order $m\alpha^6(m/M)$ in the hydrogen atom.

Atomic units ($\hbar=m=e=1$) are used throughout. It is assumed that $M\gg m$.

\section{Generalities}

\subsection{Nonrelativistic Hamiltonian}

The nonrelativistic Hamiltonian ($\mathbf{r} = \mathbf{r}_e\!-\!\mathbf{R}$) for the hydrogenlike two-body system may be written as follows:
\begin{equation}\label{Hamiltonian}
H_0 = \frac{\mathbf{P}^2}{2M}+\frac{\mathbf{p}_e^2}{2m}
      -\frac{Ze^2}{4\pi r} = \frac{\mathbf{p}^2}{2\mu} - \frac{Z}{r}
      = \frac{\mathbf{p}^2}{2\mu} + V,
\qquad \mathbf{p}_e = \mathbf{p} = -\mathbf{P},
\end{equation}
here $\mu$ is the reduced mass and $\mu=mM/(m+M)=m\,(1-\frac{m}{M}+\dots)$.

\subsection{The leading order relativistic corrections of order $m\alpha^4$}

The next order relativistic correction is expressed in terms of the Breit-Pauli Hamiltonian:
\begin{equation}\label{BPH}
\begin{array}{@{}l}\displaystyle
H^{(4)} = H_B + H_{ret},
\\[3mm]\displaystyle
H_B = -\frac{p^4}{8m^3}+\frac{[\Delta V]}{8m^2},
\\[3mm]\displaystyle
H_{ret} = -\frac{Z}{mM}\left(p^i\mathcal{W}^{ij}p^j\right).
\end{array}
\end{equation}
We include here only terms which are required for calculation of the spin-averaged energy. Here we use the notation
\begin{equation}
\mathcal{W}^{ij} = \frac{1}{2}\left(\frac{\delta^{ij}}{r}+\frac{r^ir^j}{r^3}\right).
\end{equation}

Eq.~(\ref{BPH}) will be used to write down the second-order contribution to the $m\alpha^6(m/M)$ order correction.

\subsection{Interactions at $m\alpha^6$ order}

\begin{figure}
\includegraphics[width=120mm]{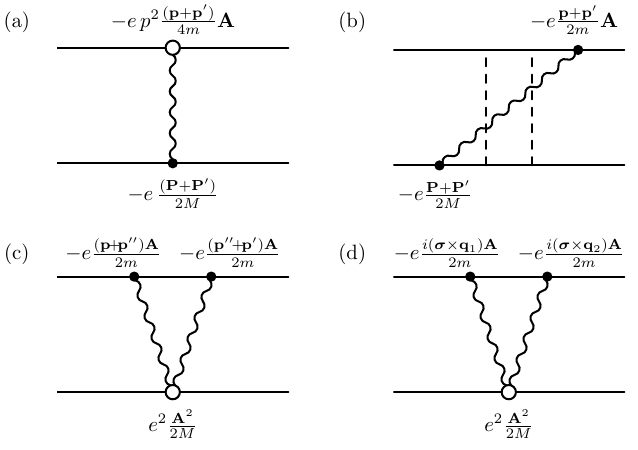}
\caption{NRQED diagrams for the recoil interactions at order $m\alpha^6(m/M)$.}\label{NRQED_diagrams}
\end{figure}

The effective Hamiltonian for the relativistic recoil is a sum of four contributions determined by the NRQED diagrams shown in Fig.~\ref{NRQED_diagrams}:
\begin{equation}\label{H6}
H^{(6)}_{rec} = H_a+H_b+H_c+H_d.
\end{equation}
Among them are the relativistic correction to the transverse photon dipole vertex $H_a$ [Fig.~\ref{NRQED_diagrams}(a)]; the retarded transverse photon exchange $H_b$ [Fig.~\ref{NRQED_diagrams}(b)]; the double transverse photon exchange with the seagull vertex on the nucleus, $\frac{Z^2}{2M}\mathbf{A}^2$, and the dipole, $-\frac{1}{m}\mathbf{p}\mathbf{A}$, and spin-orbit, $-\frac{1}{2m}\boldmath{\sigma}_e\mathbf{B}$, vertices on the electron line, Hamiltonians $H_c$ and $H_d$ [Figs.~\ref{NRQED_diagrams}(c) and \ref{NRQED_diagrams}(d)], respectively:
\begin{equation}\label{H6}
\begin{array}{@{}l}
\displaystyle
H_a =
   \frac{Z}{8m^2}
   \left\{
      p^2,\frac{p^i}{m}\left(\frac{\delta^{ij}}{r}+\frac{r^ir^j}{r^3}\right)\frac{p^j}{M}
   \right\},
\\[3mm]\displaystyle
H_b =
   \frac{Z}{8mM}\left(\Bigl[V,p^j\Bigr]\frac{r^ir^j\!-\!3\delta^{ij}r^2}{r}
      +p^j\left[\frac{p^2}{2\mu},\frac{r^ir^j\!-\!3\delta^{ij}r^2}{r}\right]\right)\Bigl[V,p^i\Bigr],
\\[3mm]\displaystyle
H_c =
   \frac{Z^2}{8m^2M}
      \left(\mathbf{p}\frac{1}{r^2}\mathbf{p}+3\frac{(\mathbf{pr})(\mathbf{rp})}{r^4}\right),
\\[3mm]\displaystyle
H_d =
   \frac{1}{4m^2M}\>\frac{Z^2}{r^4}\>.
\end{array}
\end{equation}
The sum $H^{(6)}_{rec}$ exactly matches the Hamiltonian from Ref.~\cite{YPP16}, Eq.~(D22), except for the $\delta$-function term $\frac{Z^3}{2}\pi\delta(\mathbf{r})$.

The second-order contribution to this order is
\begin{equation}
E_{rec}^{(6)} =
2\alpha^4\left\langle
   H_B\,Q (E_0-H_0)^{-1} Q\,H_{ret}
\right\rangle.
\end{equation}
It is divergent and it should be rewritten as follows:
\begin{equation}\label{2nd-order_reg}
\begin{array}{@{}l}\displaystyle
2\left\langle
   H_B\,Q (E_0-H_0)^{-1} Q\,H_{ret}
\right\rangle =
   2\left\langle
      H'_B\,Q (E_0-H_0)^{-1} Q\,H_{ret}
   \right\rangle
   +2c\left\langle VH_{ret}^{}\right\rangle
   -2c\left\langle V \right\rangle \left\langle H_{ret}^{} \right\rangle
\\[3mm]\displaystyle\hspace{25mm}
 = 2\left\langle
      H'_B\,Q (E_0-H_0)^{-1} Q\,H_{ret}
   \right\rangle
   -\frac{1}{2}\left\langle VH_{ret}^{}\right\rangle
   +\frac{1}{2}\left\langle V \right\rangle \left\langle H_{ret}^{} \right\rangle,
\end{array}
\end{equation}
where
\begin{equation}\label{HPP}
H'_{B}=H_B^{}-c(E_0\!-\!H_0)V-cV(E_0\!-\!H_0),
\qquad
c = -\frac{\mu(2\mu\!-\!m)}{4m^3}\approx-\frac{1}{4m}\,.
\end{equation}
Only the second term in the last line of Eq.~(\ref{2nd-order_reg}) diverges. It can be viewed as a new term to be added to the effective Hamiltonian (\ref{H6}).

\section{Separation of singular part}

Before starting this section we want to introduce the following notation:
\[
[p^2]_\mu = p^2\!+\!2\mu V.
\]
It is convenient for our derivations, since $[p^2]_\mu\psi_0$ is a regular function. Here $\psi_0$ is the solution of the stationary Schr\"odinger equation,
\begin{equation}\label{Schrodinger}
(H_0-E_0)\psi_0=0.
\end{equation}

Now let us consider the operators from the effective Hamiltonian in more detail.

\subsection{Relativistic correction to the transverse photon dipole vertex, $H_a$}

The relativistic correction to the transverse photon dipole vertex can be written as
\begin{equation}
H_a =
   \frac{Z}{4m^3M}
   \left\{
   p^2,\left[
      p^i\mathcal{W}^{ij}p^j
   \right]
   \right\} =
   \frac{Z^2\mu}{2m^2}\left\{ V,H_{ret} \right\}
   -\frac{1}{4m^2}\left\{ [p^2]_\mu, H_{ret} \right\}.
\end{equation}
We want to note that the second term is finite. Curly brackets denote: $\{A,B\}=AB+B^*\!A^*$. Then
\[
\begin{array}{@{}l}\displaystyle
V H_{ret} =
   -\frac{Z}{2mM}\:
   V\left[
      p^i\left(\frac{\delta^{ij}}{r}+\frac{r^ir^j}{r^3}\right)p^j
   \right] =
%   -\frac{1}{2mM}\:
%   \frac{1}{r}\left[
%      \nabla^i\left(\frac{\delta^{ij}}{r}+\frac{r^ir^j}{r^3}\right)\nabla^j
%   \right] =
\\[3mm]\displaystyle\hspace{14mm}
 = \frac{Z^2}{2mM}
   \left[\nabla^i,\frac{1}{r}\right]
      \left(\frac{\delta^{ij}}{r}+\frac{r^ir^j}{r^3}\right)\nabla^j
   -\frac{Z^2}{2mM}
   \left[
      \nabla^i\left(\frac{\delta^{ij}}{r^2}+\frac{r^ir^j}{r^4}\right)\nabla^j
   \right]
\\[3mm]\displaystyle\hspace{14mm}
% = -\frac{1}{2mM}\:
%   \frac{r^i}{r^3}
%      \left(\frac{\delta^{ij}}{r}+\frac{r^ir^j}{r^3}\right)\nabla^j
%   -\frac{1}{2mM}
%   \left[
%      \nabla^i\left(\frac{\delta^{ij}}{r^2}+\frac{r^ir^j}{r^4}\right)\nabla^j
%   \right]
%\\[3mm]\displaystyle\hspace{14mm}
 = -\frac{Z^2}{mM}\:
      \frac{1}{r^4}\left(r^i\nabla^i\right)
   +\frac{Z^2}{2mM}
   \left[
      p^i\left(\frac{\delta^{ij}}{r^2}+\frac{r^ir^j}{r^4}\right)p^j
   \right].
\end{array}
\]
Again the second term is finite. The first term may be rewritten \cite{Korobov25} as follows:
\[
\left\langle
   \frac{Z^2}{r^4}\left(r^i\nabla^i\right)
\right\rangle =
\left\langle V\boldsymbol{\mathcal{E}}\boldsymbol{\nabla} \right\rangle =
   \frac{1}{2}\left[
      4\pi\left\langle V\rho \right\rangle
      +\left\langle \boldsymbol{\mathcal{E}}^2 \right\rangle
   \right],
\]
where $\boldsymbol{\mathcal{E}}=-\boldsymbol{\nabla}V$ and $\Delta V=4\pi\rho$.
The final expression for this contribution is
\begin{equation}
\bigl\langle H_a \bigr\rangle =
   -\frac{\mu}{2m^3M}\left[
      4\pi\left\langle V\rho \right\rangle
      +\left\langle \boldsymbol{\mathcal{E}}^2 \right\rangle
   \right]
   +\frac{Z^2\mu}{2m^3M}
   \left\langle
   \left[
      p^i\left(\frac{\delta^{ij}}{r^2}+\frac{r^ir^j}{r^4}\right)p^j
   \right]
   \right\rangle
   -\frac{Z}{m^2}\left\langle [p^2]_\mu H_{ret} \right\rangle.
\end{equation}
It should be compared with Eq.~(D23) from Ref.~\cite{YPP16}.

\subsection{Retarded transverse photon exchange, $H_b$}

The retarded transverse photon exchange can be expressed as
\begin{equation}
H_b = \frac{Z}{8mM}\Bigl[V,p^i\Bigr]
   \left(\Bigl[V,p^j\Bigr]\frac{r^ir^j\!-\!3\delta^{ij}r^2}{r}+\left[\frac{p^2}{2\mu},\frac{r^ir^j\!-\!3\delta^{ij}r^2}{r}\right]p^j\right)
\end{equation}
It may be split into two parts.

The first part is expressed as
\begin{equation}\label{Hb1}
H_{b,1} = \frac{Z}{8mM}\Bigl[V,p^i\Bigr]\frac{r^ir^j\!-\!3\delta^{ij}r^2}{r}\Bigl[V,p^j\Bigr] =
   -\frac{Z^3}{8mM}\frac{r^i}{r^3}\;\frac{r^ir^j\!-\!3\delta^{ij}r^2}{r}\;\frac{r^j}{r^3} = \frac{1}{4mM}\frac{Z^3}{r^3}.
\end{equation}
Formally, it includes two regularized potentials: $V_{reg}$, and $\left[\frac{r^ir^j\!-\!3\delta^{ij}r^2}{r}\right]_{reg}$, and Eq.~(\ref{Hb1}) may not be equivalent to $V^3_{reg}$ at $r=0$. In the dimensional regularization, see Eq.~(A26) of Ref.~\cite{YPP16}, one gets the following identity for these operators:
\begin{equation}\label{YPP-A26}
\left\langle\frac{Z}{8}\Bigl[V_\epsilon,p^i\Bigr]\left[\frac{r^ir^j\!-\!3\delta^{ij}r^2}{r}\right]_\epsilon\Bigl[V_\epsilon,p^j\Bigr]\right\rangle_\epsilon =
   -\frac{1}{4}\left\langle V^3_\epsilon \right\rangle_\epsilon - \frac{7Z^3}{4}\pi\langle \delta(\mathbf{r}) \rangle.
\end{equation}
Here $\langle\cdot\rangle_\epsilon$ means that we use the dimensionally regularized solution $\psi_\epsilon$ when evaluating the expection value of the operator.

The second part is expressed as
\begin{equation}
H_{b,2} = \frac{Z}{8mM}\Bigl[V,p^i\Bigr]
   \left[\frac{p^2}{2\mu},\frac{r^ir^j\!-\!3\delta^{ij}r^2}{r}\right]p^j
\end{equation}
then using the commutator
\[
\left[\frac{p^2}{2},\frac{r^ir^j-3r^2\delta^{ij}}{r}\right] =
   2\frac{r^ir^j}{r^3}+2\frac{\delta^{ij}}{r}-i\frac{r^ip^j+r^jp^i}{r}+i\frac{r^ir^j(\mathbf{rp})}{r^3}
   +3i\frac{\delta^{ij}(\mathbf{rp})}{r}
\]
we get
\begin{equation}
H_{b,2} = \frac{Z^2}{4\mu mM}\frac{1}{r^4}(r^i\nabla^i)
   +\frac{Z^2}{8\mu mM}\left(p^i\left(\frac{\delta^{ij}}{r^2}-3\frac{r^ir^j}{r^4}\right)p^j\right).
\end{equation}

Summing up, one gets
\begin{equation}
H_b =
   \frac{1}{4mM}\frac{Z^3}{r^3}
   +\frac{1}{8\mu mM}\left[
         4\pi V\rho + \boldsymbol{\mathcal{E}}^2
      \right]
   +\frac{Z^2}{8\mu mM}\left(p^i\left(\frac{\delta^{ij}}{r^2}-3\frac{r^ir^j}{r^4}\right)p^j\right).
\end{equation}
Except for the $\delta$-function term, this exactly matches Eq.~(D24) from Ref.~\cite{YPP16}.

At this point we want to make a comment and suggestion. As seen from the above discussion, the final expressions should depend on the choice of the regularization. On the other hand, this only changes the part that depends on distributions of the $\delta$-function type or, in other words, distributions which are equal to 0 outside the domain $r=0$. Thus, we can ignore for a while the $\delta$-function type contributions in these formulas and finally reconstruct them by comparing with the known result for the $1S$ state of the hydrogen atom.

\subsection{Two-photon exchange with the seagull vertex, $H_c$ and $H_d$}

Here $H_c$ is finite and $H_d$ is singular, the sum is
\begin{equation}
H_{c\text{-}d} =
   \frac{1}{4m^2M}\,\boldsymbol{\mathcal{E}}^2
   +\frac{Z^2}{8m^2M}\left(p^i\left(\frac{\delta^{ij}}{r^2}+3\frac{r^ir^j}{r^4}\right)p^j\right).
\end{equation}

\subsection{Second-order contribution. Modification to the effective Hamiltonian}

The next contribution comes from the Breit-Pauli Hamiltonian (\ref{BPH}) as the second-order perturbation term
\begin{equation}\label{second-order_reg}
\Delta E_{2nd-order} =
\alpha^4\Bigl[
   2\left\langle H'_B\,Q (E_0-H_0)^{-1} Q\,H_{ret} \right\rangle
   +2c\left\langle VH_{ret}^{}\right\rangle
   -2c\left\langle V \right\rangle \left\langle H_{ret}^{} \right\rangle
\Bigr],
\end{equation}
and the second term is expressed
\begin{equation}
H^{(6m)}_{rec} = c\left\{V,H_{ret}\right\} =
   -\frac{c}{mM}\left[4\pi V\rho + \boldsymbol{\mathcal{E}}^2 \right]
   +\frac{Z^2c}{mM} \left[p^i\left(\frac{\delta^{ij}}{r^2}+\frac{r^ir^j}{r^4}\right)p^j\right].
\end{equation}

\subsection{The total recoil contribution of order $m\alpha^6(m/M)$}

We assume here that $m=\mu=1$. Then the final expression becomes
\begin{equation}\label{H_recoil}
\begin{array}{@{}l}\displaystyle
H_{a\text{-}d}+H^{(6m)}_{rec} =
   -\frac{\boldsymbol{\mathcal{E}}^2}{2M}
   +\frac{Z^2}{2M}\left(\mathbf{p}\frac{1}{r^2}\mathbf{p}+\frac{(\mathbf{pr})(\mathbf{rp})}{r^4}\right)
   -\frac{1}{2}\Bigl\{[p^2]_\mu,H_{ret}\Bigr\}
\\[3mm]\displaystyle\hspace{35mm}
   +\frac{\boldsymbol{\mathcal{E}}^2}{8M}
   -\frac{V^3}{4M}
   +\frac{Z^2}{8M}\left(\mathbf{p}\frac{1}{r^2}\mathbf{p}-3\frac{(\mathbf{pr})(\mathbf{rp})}{r^4}\right)
\\[3mm]\displaystyle\hspace{35mm}
   +\frac{\boldsymbol{\mathcal{E}}^2}{4M}
   +\frac{Z^2}{8M}\left(\mathbf{p}\frac{1}{r^2}\mathbf{p}+3\frac{(\mathbf{pr})(\mathbf{rp})}{r^4}\right)
\\[3mm]\displaystyle\hspace{35mm}
   +\frac{\boldsymbol{\mathcal{E}}^2}{4M}
   -\frac{Z^2}{4M} \left(\mathbf{p}\frac{1}{r^2}\mathbf{p}+\frac{(\mathbf{pr})(\mathbf{rp})}{r^4}\right)
\\[3mm]\displaystyle\hspace{9mm}
 = \frac{\boldsymbol{\mathcal{E}}^2}{8M}
   -\frac{V^3}{4M}
   +\frac{Z^2}{4M}\:\mathbf{p}\frac{1}{r^2}\mathbf{p}
   +\frac{Z^2}{4M}\left(\mathbf{p}\frac{1}{r^2}\mathbf{p}+\frac{(\mathbf{pr})(\mathbf{rp})}{r^4}\right)
   -\frac{1}{2}\Bigl\{[p^2]_\mu,H_{ret}\Bigr\}
   +C\,\frac{Z^3}{M}\,\pi\delta(\mathbf{r}).\hspace*{-10mm}
\end{array}
\end{equation}
In the last line we have introduced a still unknown term determined by the $\delta$-function distribution. The coefficient $C$ has to be identified. The total contribution to the energy is
\begin{equation}\label{Erec-final}
\Delta E^{(6)}_{rec} = \alpha^4\bigl\langle H_{a\text{-}d}+H^{(6m)}_{rec}+H_H \bigr\rangle
   +2\alpha^4\left\langle H'_B\,Q (E_0-H_0)^{-1} Q\,H_{ret} \right\rangle
   +\frac{\alpha^4}{2}\left\langle V \right\rangle \left\langle H_{ret}^{} \right\rangle,
\end{equation}
where
\[
H_H = C\,\frac{Z^3}{M}\,\pi\delta(\mathbf{r}).
\]
So far we never used that $\psi_0$ is the wave function for the hydrogen bound state, thus essentially the same derivation may be used in the case of the hydrogen molecular ion as well.

\section{Comparison with the $1S$ state result and determination of $C$}

Now we may find the coefficient $C$. For that we need the expectation values for various operators for the $nS$ states of the hydrogen atom:
\begin{equation}
\begin{array}{@{}l}\displaystyle
\left\langle p^i\,\mathcal{W}^{ij}p^j\right\rangle =
	\frac{Z^3}{n^3}\left(2-\frac{1}{n}\right),
\qquad
\left\langle\mathbf{p}\,\frac{1}{r^2}\,\mathbf{p}\right\rangle =
\left\langle p^i\frac{\mathcal{W}^{ij}}{r}p^j\right\rangle =
	\frac{Z^4}{n^3}\,\left[\frac{8}{3}-\frac{2}{3n^2}\right],
\\[4mm]\displaystyle
2\left\langle H'_B\,Q (E_0\!-\!H_0)^{-1} Q\,H_{ret}\right\rangle =
	-\frac{m^2(Z\alpha)^6}{Mn^3}\left(\frac{11}{3}+\frac{3}{n}-\frac{37}{6n^2}+\frac{2}{n^3}\right).
\end{array}
\end{equation}

Using the regularized form for the two divergent integrals:
\begin{equation}\label{cut-off_reg}
\begin{array}{@{}l}
\displaystyle
\left\langle \frac{1}{r^3} \right\rangle_{\!s} =
   \lim_{r_0\to0}
   \left\{
   \left\langle
      \frac{1}{r^3}
   \right\rangle_{\!\!r_0}\!
   +\left(\ln{r_0}\!+\!\gamma_E\right)\left\langle 4\pi\delta(\mathbf{r}) \right\rangle
   \right\} =
\\[3mm]\displaystyle\hspace{40mm}
 = -4\pi\langle\delta(\mathbf{r})\rangle
   \left[
      \ln(2Z)+\psi(n)-\psi(1)-\ln{n}
      -\frac{1}{2}+\frac{1}{2n}
   \right],
\\[3mm]\displaystyle
\left\langle \frac{1}{r^4} \right\rangle_{\!s} =
   \lim_{r_0\to0}
   \left\{
   \left\langle
      \frac{1}{r^4}
   \right\rangle_{\!\!r_0}\!
   -\left[
      \frac{1}{r_0}\left\langle 4\pi\delta(\mathbf{r}) \right\rangle
      +\left(\ln{r_0}\!+\!\gamma_E\right)\left\langle 4\pi\delta'(\mathbf{r}) \right\rangle
   \right]
   \right\} =
\\[3mm]\displaystyle\hspace{40mm}
 = 8\pi\langle\delta(\mathbf{r})\rangle
   \left[
      \ln(2Z)+\psi(n)-\psi(1)-\ln{n}
      -\frac{5}{3}+\frac{1}{2n}+\frac{1}{6n^2}
   \right].
\end{array}
\end{equation}
we get the final expression for the energy shift due to the recoil contribution:
\[
\Delta E^{(6)}_{rec} =
   \frac{4Z^6\alpha^4}{Mn^3}\left(-\frac{3}{8}+\frac{3}{8 n}-\frac{1}{n^2}+\frac{1}{2 n^3}\right)
\]
That should be compared with the ground state result for the pure recoil correction \cite{Pachucki95}:
\begin{equation}
\Delta E_{rec}^{(6)}(1S) =
   \frac{Z^6\alpha^4}{M}\left(4\ln{2}-\frac{7}{2}\right).
\end{equation}
This gives the value for the unknown coefficient $C_s = 4\ln{2}-3$, where the subscript $s$ stands for the coordinate cut-off regularization \cite{Korobov25}.
In case of the dimensional regularization we get: $C_\epsilon =  4\ln{2}-9/2$, and then our expression (\ref{H_recoil}) and (\ref{Erec-final}) exactly match Eq.~(D26) from Ref.~\cite{YPP16}.

\section{Recoil corrections for the hydrogen molecular ion.}

\subsection{Nonrelativistic Hamiltonian}

The nonrelativistic Hamiltonian for the hydrogen molecular ion in the center-of-mass frame [here $\mathbf{r}_a = \mathbf{r}_e\!-\!\mathbf{R}_a$ for $a=(1,2)$] is defined as
\begin{equation}\label{Hamiltonian}
H_0 = \frac{\mathbf{P}_1^2}{2M_1}+\frac{\mathbf{P}_2^2}{2M_2}+\frac{\mathbf{p}_e^2}{2m}
      -\frac{Z_1}{r_1}-\frac{Z_2}{r_2}+\frac{Z_1Z_2}{R}
 = \frac{\mathbf{p}_1^2}{2\mu_{1}} + \frac{\mathbf{p}_2^2}{2\mu_2}+\frac{\mathbf{p}_1\mathbf{p}_2}{m} + V_1+V_2+V_{12},
\end{equation}
where $\mu_1^{-1}=M_1^{-1}\!+\!m^{-1}$ and $\mu_2^{-1}=M_2^{-1}\!+\!m^{-1}$ are reduced masses of two nuclei, $\mathbf{p}_e=\mathbf{p}_1\!+\!\mathbf{p}_2$, $\mathbf{P}_1=-\mathbf{p}_1$, and $\mathbf{P}_2=-\mathbf{p}_2$. We also use the following notation $V_e=V_1\!+\!V_2$ for the electron potential energy.

The part of the Breit-Pauli Hamiltonian, required for the spin-averaged recoil correction to the energy, now is expressed as
\begin{equation}
\begin{array}{@{}l}\displaystyle
H_B = -\frac{p^4_e}{8m^3}+\frac{[\Delta V_e]}{8m^2},
\\[3mm]\displaystyle
H_{ret} = \frac{Z_1}{mM_1}\left(p_e^i\mathcal{W}_1^{ij}P_1^j\right)+\frac{Z_2}{mM_2}\left(p_e^i\mathcal{W}_2^{ij}P_2^j\right),
\\[3mm]\displaystyle
H_{so} =
   \frac{Z_1}{2m^2}\,\frac{[\mathbf{r}_1\!\times\!\mathbf{p}_e]}{r_1^3}\,\mathbf{s}_e
   -\frac{Z_1}{mM_1}\,\frac{[\mathbf{r}_1\!\times\!\mathbf{P}_1]}{r_1^3}\,\mathbf{s}_e
   +\frac{Z_2}{2m^2}\,\frac{[\mathbf{r}_2\!\times\!\mathbf{p}_e]}{r_2^3}\,\mathbf{s}_e
   -\frac{Z_2}{mM_2}\,\frac{[\mathbf{r}_2\!\times\!\mathbf{P}_2]}{r_2^3}\,\mathbf{s}_e.
\end{array}
\end{equation}

We use as well the following notation:
\begin{equation}
[p^2]_\mu = p^2_e+2V_\mu,
\qquad
V_\mu = \mu_1 V_1+\mu_2V_2.
\end{equation}
and it is easy to see that $[p^2]_\mu\psi_0$ is a regular function, when $\psi_0$ is a solution of the nonrelativistic Schr\"odinger equation.

\subsection{The effective Hamiltonian in $m\alpha^6(m/M)$ order}

The recoil effective Hamiltonian is \cite{Korobov20}
\begin{equation} \label{hrec}
\begin{array}{@{}l}\displaystyle
H_{rec}^{(6)} = H_a + H_b + H_c + H_d +H_H,
\\[3mm]\displaystyle
H_a = -\frac{1}{4m^2}\left\{p_e^2,H_{ret}\right\} =
   \frac{Z^2\mu}{2m^2}\left\{ V_\mu,H_{ret} \right\}
   -\frac{1}{4m^2}\left\{ [p^2]_\mu, H_{ret} \right\},
\\[3mm]\displaystyle
H_b =
   \frac{1}{4m}
      \left[
         \frac{Z_1^3}{M_1}\frac{1}{r_1^3}+\frac{Z_2^3}{M_2}\frac{1}{r_2^3}
         +\frac{Z_1^2Z_2}{M_1}\frac{(\mathbf{r}_1\mathbf{r}_2)}{r_1^2r_2^3}
         +\frac{Z_1Z_2^2}{M_2}\frac{(\mathbf{r}_1\mathbf{r}_2)}{r_1^3r_2^2}
      \right]
\\[3mm]\hspace{8mm}\displaystyle
      +\frac{Z_1^2}{8m^2M_1}
         \left[
            \frac{1}{r_1^4}+\mathbf{p}_e\frac{1}{r_1^2}\mathbf{p}_e
               -3\frac{(\mathbf{p}_e\mathbf{r}_1)(\mathbf{r}_1\mathbf{p}_e)}{r_1^4}
         \right]
      +\frac{Z_2^2}{8m^2M_2}
         \left[
            \frac{1}{r_2^4}+\mathbf{p}_e\frac{1}{r_2^2}\mathbf{p}_e
               -3\frac{(\mathbf{p}_e\mathbf{r}_2)(\mathbf{r}_2\mathbf{p}_e)}{r_2^4}
         \right],
\\[3mm]\displaystyle
H_c =
   \frac{Z_1^2}{8m^2M_1}
   \left[
      \mathbf{p}_e\frac{1}{r_1^2}\mathbf{p}_e+3\frac{(\mathbf{p}_e\mathbf{r}_1)(\mathbf{r}_1\mathbf{p}_e)}{r_1^4}
   \right]
   +\frac{Z_2^2}{8m^2M_2}
   \left[
      \mathbf{p}_e\frac{1}{r_2^2}\mathbf{p}_e+3\frac{(\mathbf{p}_e\mathbf{r}_2)(\mathbf{r}_2\mathbf{p}_e)}{r_2^4}
   \right],
\\[3mm]\displaystyle
H_d =
   \frac{Z_1^2}{4m^2M_1}\>\frac{1}{r_1^4}
   +\frac{Z_2^2}{4m^2M_2}\>\frac{1}{r_2^4}\>.
\end{array}
\end{equation}

The explicit expression for the $\{V_\mu, H_{ret}\}$ operator ($\mu=1$) may be written as
\begin{equation}
\begin{array}{@{}l}\displaystyle
\left\langle V_\mu H_{ret}\right\rangle = \left\langle V_1 H_{ret,1}\right\rangle+\left\langle V_2 H_{ret,2}\right\rangle
   +\left\langle V_1 H_{ret,2}\right\rangle+\left\langle V_2 H_{ret,1}\right\rangle,
\\[3mm]\displaystyle
\left\langle V_1 H_{ret,1}\right\rangle =
   -\frac{1}{2M}\langle\boldsymbol{\mathcal{E}}^2\rangle
   -\frac{Z^2}{2M} \left\langle p_e^i\frac{W_1^{ij}}{r_1}P_1^j\right\rangle.
\end{array}
\end{equation}
The cross terms are finite and can thus be calculated numerically. We recall here that we ignore the $\delta$-function terms in the intermediate formulas.

The effective Hamiltonian~(\ref{hrec}) is incomplete, because it does not include contributions from the hard photon energy region. We use the regularized form for the singular operators as in Eq.~(\ref{cut-off_reg}), and thus the high-energy Hamiltonian should be chosen in the form
\begin{equation}\label{H_H}
H_H = \Bigl(4\ln{2}-3\Bigr)\frac{Z^3}{M}\,\Bigl(\pi\delta(\mathbf{r}_1)+\pi\delta(\mathbf{r}_2)\Bigr).
\end{equation}

Then the second-order contribution is expressed
\begin{equation}
\begin{array}{@{}l}\displaystyle
\Delta E_{rec}^{2^{nd}-order} = \Delta E_{ret} + \Delta E_{so\hbox{-}so}^{(0)},
\\[3mm]\displaystyle
\Delta E_{ret} =
2\alpha^2\left\langle
   H_B\,Q (E_0-H_0)^{-1} Q\,H_{ret}
\right\rangle,
\\[3mm]\displaystyle
\Delta E_{so\hbox{-}so}^{(0)} =
   \alpha^2\left\langle
      H_{so} Q (E_0-H_0)^{-1} Q H_{so}
   \right\rangle^{(0)}.
\end{array}
\end{equation}
The last one is the spin-orbit scalar contribution, since $H_{so}$ is a vector operator, the second-order term has contributions of rank 0, 1, and 2. Its mean value is finite.

\subsection{Modifications to the effective Hamiltonian resulting from the second-order contributions}

Since $\Delta E_{ret}$ is divergent, it should be transformed as follows:
\begin{equation}
\begin{array}{@{}l}\displaystyle
\left\langle
   H_B\,Q (E_0-H_0)^{-1} Q\,H_{ret}
\right\rangle =
   \left\langle
      H'_B\,Q (E_0-H_0)^{-1} Q\,H_{ret}
   \right\rangle
   -\frac{1}{2}\left\langle V_eH_{ret}^{} \right\rangle
   +\frac{1}{2}\left\langle V_e \right\rangle \left\langle H_{ret}^{} \right\rangle
\end{array}
\end{equation}
Additional term, which should be added to the effective Hamiltonian, has the form
\begin{equation}\label{H_mret}
\begin{array}{@{}l}
\displaystyle
H_{rec}^{(6m)} =
   -\frac{1}{4}\left\{ V_e,H_{ret}^{} \right\}.
\end{array}
\end{equation}

\subsection{Recoil energy shift via finite operators}

Now we are ready to express the recoil correction in terms of the finite-value operators.

The total sum of the effective Hamiltonian $H_{rec}^{(6)}$ and $H_{rec}^{(6m)}$ is expressed ($m=1$):
\begin{equation}\label{final-effHam}
\begin{array}{@{}l}\displaystyle
\left\langle H_{rec}^{(6)}+H^{(6m)}_{rec}\right\rangle =
   \frac{1}{8}
      \left\langle
         \frac{Z_1^2}{M_1}\:\frac{1}{r_1^4}
         +\frac{Z_2^2}{M_2}\:\frac{1}{r_2^4}
      \right\rangle
   +\frac{1}{4}
      \left\langle
         \frac{Z_1^3}{M_1}\frac{1}{r_1^3}+\frac{Z_2^3}{M_2}\frac{1}{r_2^3}
      \right\rangle
\\[3mm]\hspace{20mm}\displaystyle
   -\frac{1}{2}\left\langle [p^2]_\mu H_{ret} \right\rangle
   +\frac{1}{2}\left\langle V_1 H_{ret,2} \right\rangle
   +\frac{1}{2}\left\langle V_2 H_{ret,1} \right\rangle
\\[3mm]\hspace{20mm}\displaystyle
   +\frac{1}{4}
      \left\langle
         \frac{Z_1^2Z_2}{M_1}\frac{(\mathbf{r}_1\mathbf{r}_2)}{r_1^2r_2^3}
         +\frac{Z_1Z_2^2}{M_2}\frac{(\mathbf{r}_1\mathbf{r}_2)}{r_1^3r_2^2}
      \right\rangle
\\[3mm]\hspace{20mm}\displaystyle
      +\frac{1}{4}
         \left\langle
            \frac{Z_1^2}{M_1}\,\mathbf{p}_e\frac{1}{r_1^2}\mathbf{p}_e
            +\frac{Z_2^2}{M_2}\,\mathbf{p}_e\frac{1}{r_2^2}\mathbf{p}_e
         \right\rangle
      -\frac{1}{4}
         \left\langle
            \frac{Z_1^2}{M_1}\,\mathbf{p}_e\frac{W_1^{ij}}{r_1}\mathbf{P}_1
            +\frac{Z_2^2}{M_2}\,\mathbf{p}_e\frac{W_2^{ij}}{r_2}\mathbf{P}_2
         \right\rangle.
\end{array}
\end{equation}

Then we replace the expectation values for the diverent operators $1/r^3$ and $1/r^4$ by the regularized integrals, as in Eq.~(\ref{cut-off_reg}) and this makes our evaluation of the effective Hamiltonian (\ref{final-effHam}) finite. Thus the final expression for the recoil energy shift becomes
\begin{equation}\label{final-shift}
\Delta E^{(6)}_{rec} = \alpha^4\bigl\langle H_{rec}^{(6)}+H^{(6m)}_{rec} +H_H\bigr\rangle
   +2\alpha^4\left\langle H'_B\,Q (E_0\!-\!H_0)^{-1} Q\,H_{ret} \right\rangle
   +\frac{\alpha^4}{2}\left\langle V_e \right\rangle \left\langle H_{ret}^{} \right\rangle
   +\Delta E_{so\hbox{-}so}^{(0)}.
\end{equation}
All the operators in the above expression have finite expectation values if the wave function is a solution of the nonrelativistic Schr\"odinger equation for some bound state and can now be calculated numerically.

Another way to achieve these results is to use the dimensional regularization. It is important to note that all the cross-terms of the effective Hamiltonian have finite expectation values and do not require regularization. Thus we may restrict consideration to the pairwise interactions only. Then using the derivation of the NRQED interaction potentials presented in the Appendix we obtain expressions similar to Eq.(\ref{final-effHam}) and $H_H$ as in Eq.~(\ref{H_H_dim}). Then taking the identities \cite{Korobov25}:
\begin{equation}
\begin{array}{@{}l}\displaystyle
\left\langle V^3_\epsilon\right\rangle_{\!\epsilon} =
   -\left\langle \frac{Z^3}{r^3}\right\rangle_{\!s}+Z^3\langle 4\pi\delta(\mathbf{r})\rangle\left(-\frac{1}{4\epsilon}-\frac{3\ln{4\pi}}{4}+\frac{\gamma_E}{4}\right),
\\[3mm]\displaystyle
\left\langle \boldsymbol{\mathcal{E}}_{\epsilon}^2 \right\rangle_{\!\epsilon} =
   \left\langle \frac{Z^2}{r^4}\right\rangle_{\!s}
   +2Z^3 \left\langle 4\pi\delta(\mathbf{r})\right\rangle
   \left[-\frac{1}{4\epsilon}-\frac{3\ln{4\pi}}{4}+\frac{\gamma_E}{4}+\frac{3}{2}\right],
\end{array}
\end{equation}
or identities from Ref.~\cite{YPP16}, Eq. (A22) and (A24), one should arrive exactly at Eqs.~(\ref{final-effHam}) and (\ref{final-shift}).

\section{Summary}

The set of finite operators in Eqs.~(\ref{final-effHam}) and (\ref{final-shift}), which allows one to numerically calculate the recoil correction of the order of $m(Z\alpha)^6(m/M)$ for three-particle systems similar to the molecular hydrogen ion, is the main result of the present work. The only requirement for obtaining the correct answer is that we should use suitably regularized divergent integrals as defined in Eq.~(\ref{cut-off_reg}).

As shown in our previous work \cite{Korobov25}, the combination of distributions
\begin{equation}
   \left\langle
     \boldsymbol{\mathcal{E}}^2
   \right\rangle
   -2\mu\left\langle V^3 \right\rangle + \left\langle V(4\pi\rho) \right\rangle
\end{equation}
is finite, no matter which regularization scheme is used. That makes dimensional regularization preferable for rigorous derivation of the recoil, since in this case $\langle V(4\pi\rho) \rangle\equiv0$. However, it is more convenient to express and calculate the divergent matrix elements $\langle [1/r^3]_\epsilon\rangle$ and $\langle [1/r^4]_\epsilon\rangle$ in terms of $\langle 1/r^3 \rangle_s$ and $\langle 1/r^4\rangle_s$ distributions, thus, we will eventually arrive at the same set of expressions.

It is useful to note that the final expression for the effective Hamiltonian (\ref{H_recoil}) or (\ref{final-effHam}) contains a combination: $\langle 1/r^4\rangle+2\langle 1/r^3 \rangle$, which can actually be replaced by the distribution $\langle V(4\pi\rho) \rangle$ and a set of finite operators. In the case of the coordinate cut-off regularization and/or mass regularization, $\langle V(4\pi\rho) \rangle$ contains a divergent part (see Ref.~\cite{Korobov25}), which is compensated by the contribution from the hard-scale energy region. For the case of mass regularization such a cancellation was demonstrated in Ref.~\cite{Pachucki95}.

We still have to make one more remark on numerical calculation of the operators appearing in Eq.~(\ref{final-effHam}). In particular, in order to compute $\langle [p^2]_\mu H_{ret}\rangle$ we can use that $[p^2]_\mu\psi_0$ is a regular function. Therefore we can split the operator and represent it as a sum:
\begin{equation}
\langle [p^2]_\mu H_{ret}\rangle =
 \sum_{n=1}^N \langle \psi_0| [p^2]_\mu |\psi_n\rangle\langle \psi_n|H_{ret}|\psi_0\rangle,
\end{equation}
where $\psi_n$ is a finite set of orthogonal functions, spanning over finite subspace of the Hilbert space of operator $H_0$. When $N$ goes to infinity, and $\psi_n$ are a complete set of functions for the Hilbert space of the problem, then the sum should converge to exact value. This is very similar to what we do when calculating second-order contributions in perturbation theory.

\section{Acknowledgements}

The author wants to acknowledge fruitful discussions with Zhen-Xiang Zhong and J.-P.~Karr.

%\newpage
\appendix

\section{Recoil operators in dimensional regularization}

Here we follow derivation of the recoil correction done in \cite{YPP16}. We recall that $d=3\!-\!2\epsilon$ and
\begin{equation}\label{singular_operators}
\begin{array}{@{}l}\displaystyle
\left[\frac{Z^3}{r^3}\right]_\epsilon =
   Z^3\pi\delta(\mathbf{r})\:\frac{1}{\epsilon}
   +U_{\rm finite},
\\[3mm]\displaystyle
\left[\frac{Z^2}{r^4}\right]_\epsilon =
   -Z^3\pi\delta(\mathbf{r})\:\frac{2}{\epsilon}
   +W_{\rm finite}.
\end{array}
\end{equation}
And we set $m=1$. Below is a list of recoil operators presented in Fig. 1, and their expressions in the dimensionally regularized QED:

(i) The relativistic correction to the transverse photon dipole vertex
\begin{equation}
\begin{array}{@{}l}\displaystyle
H_a =
   \frac{Z}{8M}
   \left\{
      p^2,p^i\left[\frac{\delta^{ij}}{r}+\frac{r^ir^j}{r^3}\right]_\epsilon p^j
   \right\}
\\[3mm]\displaystyle\hspace{15mm}
 = -\frac{1}{2M}\left[\frac{Z^2}{r^4}\right]_\epsilon
   +\frac{Z}{2M}p^i\left(E_0+\frac{Z}{r}\right)\left(\frac{\delta^{ij}}{r}+\frac{r^ir^j}{r^3}\right)p^j
   -\frac{Z^3}{M}\pi\delta(\mathbf{r}).
\end{array}
\end{equation}

(ii) Retarded transverse photon exchange
\begin{equation}
\begin{array}{@{}l}\displaystyle
H_b =
   \frac{Z}{8M}\left(\Bigl[V_\epsilon,p^j\Bigr]\left[\frac{r^ir^j\!-\!3\delta^{ij}r^2}{r}\right]_\epsilon
      +p^j\left[\frac{p^2}{2},\left[\frac{r^ir^j\!-\!3\delta^{ij}r^2}{r}\right]_\epsilon\right]\right)\Bigl[V_\epsilon,p^i\Bigr]
\\[3mm]\displaystyle\hspace{15mm}
 = -\frac{1}{8M}\left[\frac{Z^2}{r^4}\right]_\epsilon
   +\frac{1}{4M}\left[\frac{Z^3}{r^3}\right]_\epsilon
   -\frac{Z^3}{M}\pi\delta(\mathbf{r})
   +\frac{Z^2}{8M}p^i\left(\frac{\delta^{ij}}{r^2}-3\frac{r^ir^j}{r^4}\right)p^j
\end{array}
\end{equation}

(iii) The double transverse photon exchange with the dipole vertices on the electron line
\begin{equation}
H_c =
   \frac{Z^2}{8M}p^i\left(\frac{\delta^{ij}}{r^2}+3\frac{r^ir^j}{r^4}\right)p^j.
\end{equation}
This operator is finite.

(iv) The double transverse photon exchange with the spin-orbit vertices on the electron line
\begin{equation}
H_d = \frac{\sigma^{ij}\sigma^{ij}}{8dM}\left[\frac{Z^2}{r^4}\right]_\epsilon =
   \frac{d\!-\!1}{8M}\left[\frac{Z^2}{r^4}\right]_\epsilon =
   \frac{1}{4M}\left[\frac{Z^2}{r^4}\right]_\epsilon+\frac{Z^3}{2M}\pi\delta(\mathbf{r})\:.
\end{equation}

The second-order contribution is expressed by Eq.~(\ref{second-order_reg}) with $c=-1/4$, and
\begin{equation}
\begin{array}{@{}l}\displaystyle
\frac{1}{2}\left\{ V,H_{ret} \right\} =
   \frac{1}{4M}
   \left\{
   \frac{Z}{r},\left[
      p^i\left(\frac{\delta^{ij}}{r}+\frac{r^ir^j}{r^3}\right)p^j
   \right]
   \right\} =
\\[3mm]\displaystyle\hspace{15mm}
 = -\frac{1}{2M}\left[\frac{Z^2}{r^4}\right]_\epsilon
   +\frac{Z^2}{2M}p^i\left(\frac{\delta^{ij}}{r^2}+\frac{r^ir^j}{r^4}\right)p^j
   -\frac{Z^3}{M}\pi\delta(\mathbf{r}).
\end{array}
\end{equation}

The singular part is expressed by (with account of Eq.~(\ref{singular_operators}))
\begin{equation}
\frac{1}{8M}\left[\frac{Z^2}{r^4}\right]_\epsilon+\frac{1}{4M}\left[\frac{Z^3}{r^3}\right]_\epsilon
 = \frac{1}{8M}\left(2U_{\rm finite}+W_{\rm finite}\right),
\end{equation}
and regularization parameter $\epsilon$ cancels out. Then the complete expression for the energy shift for the $nS$ states of the hydrogen atom in order $m\alpha^6(m/M)$ from the soft photon energy region is
\begin{equation}
\Delta E^{(6)}_{rec} =
   \frac{4Z^6}{Mn^3}\left(\frac{1}{8}+\frac{3}{8 n}-\frac{1}{n^2}+\frac{1}{2 n^3}\right),
\end{equation}
and is zero for the hydrogen ground state.

The remaining part comes from the hard photon energy region, \cite{Melnikov02} Eq.~(23),
\begin{equation}\label{H_H_dim}
H_H=\frac{1}{M}\left(4\ln{2}-\frac{7}{2}\right)\pi\delta(\mathbf{r}).
\end{equation}


\begin{thebibliography}{99}

\bibitem{Korobov08} V.I.~Korobov,
   Relativistic corrections of $m\alpha^6$ order to the rovibrational spectrum of H$_2^+$ and HD$^+$ molecular ions.
   Phys. Rev. A~\textbf{77}, 022509 (2008).

\bibitem{Korobov21} V.I.~Korobov and J.-Ph.~Karr,
  Rovibrational spin-averaged transitions in the hydrogen molecular ions,
  Phys. Rev. A~\textbf{104}, 032806 (2021).

\bibitem{Zhong18} Z.-X.~Zhong, W.-P.~Zhou, and X.-S.~Mei.
   Spin-averaged effective Hamiltonian of orders $m\alpha^6$ and $m\alpha^6(m/M)$ for hydrogen molecular ions.
   Phys. Rev. A~\textbf{98}, 032502 (2018).

\bibitem{Salpeter52} E.E.~Salpeter,
  Mass corrections to the fine structure of hydrogen-like atoms.
  Phys.\ Rev.\ \textbf{87}, 328 (1952).

\bibitem{GrotchYennie} H.~Grotch and D.R.~Yennie,
  Effective potential model for calculating nuclear corrections to the energy levels of hydrogen.
  Rev.\ Mod.\ Phys.\ \textbf{41}, 350 (1969).

\bibitem{Shabaev85} V.M.~Shabaev, Mass recoil in a strong nuclear field.
  Theor.\ Math.\ Phys.\ \textbf{63}, 588 (1985).

\bibitem{Yelkhovsky94} A.S.~Yelkhovsky,
  Recoil Correction in the Dirac-Coulomb Problem.
  hep-th/9403095

\bibitem{Pachucki95} K.~Pachucki and H.~Grotch.
   Pure recoil corrections to the hydrogen energy levels.
   Phys.\ Rev.~A \textbf{51}, 1854 (1995).

\bibitem{Melnikov02} I.~Blockland, A.~Czarnecki, and K.~Melnikov,
  Expansion of bound-state energies in powers of $m/M$ and $(1\!-\!m/M)$.
  Phys.\ Rev.~D \textbf{65}, 073015 (2002).

\bibitem{Korobov25} V.I.~Korobov,
  Relativistic corrections of order $m\alpha^6$: singular operators and regularization.
  arXiv:2505.16645, submitted to Phy.\ Rev.~A.

\bibitem{YPP16} V.~Patk\'{o}\v{s}, V.A.~Yerokhin, K.~Pachucki.
   Higher-order recoil corrections for triplet states of the helium atom.
   Phys. Rev.~A \textbf{94}, 052508 (2016).

%\bibitem{Pac97} K.~Pachucki.
%   Effective Hamiltonian approach to the bound state: Positronium hyperfine structure.
%   Phys.\ Rev.~A \textbf{56}, 297 (1997).

\bibitem{Korobov20} V.I. Korobov, J.-Ph. Karr, M. Haidar, and Z.-X. Zhong.
   Hyperfine structure in the H$_2^+$ and HD$^+$ molecular ions at order $m\alpha^6$,
   Phys. Rev. A~\textbf{102}, 022804 (2020).

%\bibitem{Araki57} H.~Araki, Quantum-electrodynamical corrections to energy-levels of helium.
%  Prog.\ Theor.\ Phys. {\bf 17}, 619 (1957).

%\bibitem{Sucher58} J.~Sucher,
%  Energy levels of the two-electron atom to order $\alpha^3$ Ry; Ionization energy of helium.
%  Phys.Rev. {\bf 109}, 1010 (1958).

\end{thebibliography}
\end{document}